\documentclass[prl,twocolumn,showpacs,superscriptaddress,nofootinbib]{revtex4-1}

\usepackage{amsfonts}
\usepackage{amsmath}
\usepackage{amssymb}
\usepackage{bm}
\usepackage{dcolumn}
\usepackage{graphicx}
\usepackage{graphics}
\usepackage[latin1]{inputenc}
\usepackage{latexsym}
\usepackage{rotating}
\usepackage{xspace} % Sensible space treatment at end of simple macros
\usepackage[usenames]{color}
\usepackage{mathrsfs}
\usepackage{subfigure}
\usepackage{multirow}
\usepackage{ulem}
\usepackage{outlines}
\usepackage{enumitem}
\setenumerate[1]{label=\Roman*.}
\setenumerate[2]{label=\Alph*.}
\setenumerate[3]{label=\roman*.}
\setenumerate[4]{label=\alph*.}

\definecolor {darkgreen}{rgb}{0.2,0.7,0.2}

\newcommand\be{\begin{equation}}
\newcommand\ba{\begin{eqnarray}}
\newcommand\ee{\end{equation}}
\newcommand\ea{\end{eqnarray}}

\newcommand\bw{\begin{widetext}}
\newcommand\ew{\end{widetext}}

\newcommand{\nn}{\nonumber}

\begin{document}
\title{Scalar Tops and Perturbed Quadrupoles: \\Probing Fundamental Physics with Spin-Precessing Binaries}

\author{Nicholas Loutrel}
\affiliation{eXtreme Gravity Institute, Department of Physics, Montana State University, Bozeman, MT 59717, USA.}

\author{Takahiro Tanaka}
\address{Department of Physics, Kyoto University, Kyoto 606-8502, Japan}
\affiliation{Yukawa Institue for Theoretical Physics, Kyoto University, 606-8502, Kyoto, Japan}

\author{Nicol\'as Yunes}
\affiliation{eXtreme Gravity Institute, Department of Physics, Montana State University, Bozeman, MT 59717, USA.}

\date{\today}

%%%%%%%%%%%%%%%%%%%%%%%%%%%%%%%%%%%%%%%%%%%%%%%%%
\begin{abstract} 

Parity violation in the gravitational interaction has an important impact on fundamental observables and the evolution of the universe. We here investigate for the first time our ability to probe the parity violating nature of the gravitational interaction using gravitational waves from spin-precessing binaries. Focusing on dynamical Chern-Simons gravity, we derive the spin-precession equations, calculate the gravitational waves emitted by spin-precessing, quasi-circular black hole binaries and estimate the level to which the theory could be constrained with future gravitational wave observations.

\end{abstract}


% 04.30-w Gravitational Waves
% 04.30.Db Wave generation and sources
% 04.50.Kd Modified theories of gravity
% 04.25.-g Approximation methods; equations of motion
% 04.25.Nx Post-Newtonian approximation; perturbation theory; related approximations
% 97.60.Jd Neutron stars

\maketitle

%%%%%%%%%%%%%%%%%%%%%%%%%%%%%%%%%%%%%%%%%%%%%%%%%

\vspace{0.3cm}
\noindent {\emph{Introduction.}}~The detection of gravitational waves (GWs) from the coalescence of compact binaries has opened the door to precision tests of the fundamental nature of the gravitational interaction in the extreme gravity regime, where the gravitational interaction and the spacetime curvature are dynamical, nonlinear and large~\cite{TheLIGOScientific:2016src, Yunes:2016jcc}. Many modified theories of gravity predict strong deviations from General Relativity (GR) in this regime, while passing Solar System constraints. Some of these theories are constructed as a low energy limit of some high energy, effective field theory, such as string theory or loop quantum gravity. The detections of GWs by the Advanced Laser Interferometer Gravitational Wave Observatory (aLIGO) and Advanced VIRGO (aVIRGO) detectors have allowed us to place constraints on such hypothetical deviations from GR, including constraints on generic ppE parameters~\cite{TheLIGOScientific:2016src, Yunes:2016jcc}, and even constraints on the speed of the graviton~\cite{Monitor:2017mdv}.

One particular aspect of the gravitational interaction that is largely unconstrained is parity violation. Studying parity violation in GR has proven difficult due to the fact that parity operations must be made on three dimensional hypersurfaces foliating the spacetime manifold, and are thus slicing, or coordinate, dependent operations~\cite{Alexander:2017jmt}. Nevertheless, parity violation can have important physical consequences that will lead to observable signatures necessary to constrain parity violation in gravity. 

%Parity violation is intricately related to cosmological baryogenesis and the baryon asymmetry.

The simplest theory that generates parity violation in the gravitational sector is dynamical Chern-Simons (dCS) gravity~\cite{jackiw:2003:cmo, Alexander:2009tp}, which modifies the Einstein Hilbert action through the coupling of a pseudo-scalar field $\vartheta$ to the parity odd Pontryagin density ${^{\star}R} R = (1/2) \epsilon^{\mu \nu \rho \sigma} R_{\alpha \beta \rho \sigma} {R_{\mu \nu}}^{\alpha \beta}$, with $R_{\mu \nu \rho \sigma}$ the Riemann curvature tensor. While the action is parity invariant, dCS gravity is said to be parity violating, in the sense that modifications only appear in systems that are odd under spatial reflection. For example, GWs have been shown to exhibit amplitude birefringence~\cite{PhysRevLett.83.1506}, which causes left-handed (right-handed) GWs to be suppressed (enhanced) as the waves propagate. Matter and antimatter couple to right- and left-handed GWs respectively, and thus amplitude birefringence provides a natural means to explain the observed baryon asymmetry of the Universe~\cite{PhysRevLett.96.081301}.

Constraining dCS gravity and parity violation with GWs has proven to be difficult so far due to degeneracies between propagation effects and the orientation and sky angles of binary systems emitting GWs, and degeneracies between generation effects and the spin magnitudes of the components of the binary. This is why the current best constraint on dCS gravity comes from the observation of gravitomagnetic effects in frame-dragging and geodetic precession by the Gravity Probe B and LAGEOS missions, respectively. These Solar System experiments have constrained the dimensional coupling constant of the theory to $\xi^{1/4} \lesssim 10^{8} \; \text{km}$~\cite{AliHaimoud:2011fw}, one of the least stringent constraints of any modified theory of gravity that predicts deviations from GR on large scales.

%Paper with Stephon
The degeneracies between dCS corrections and spin effects, however, can be broken if the binary system precesses due to spin-orbit interactions. Most studies to date had focused on non-precessing systems, where the spins of the compact object are aligned or anti-aligned with the orbital angular momentum~\cite{Yagi:2011xp, Yagi:2012vf}. The coupling between the spin and the orbital angular momentum of the binary forces all momenta to precess about the total angular momentum of the system. Gravitational waves emitted by these binaries exhibit amplitude modulations since the emission is weakly beamed along the orbital angular momentum. This modulation encodes the spin of the orbiting bodies and could thus be used to break the degeneracies between dCS corrections and spin, just as in GR it can break degeneracies that allow one to distinguish between GWs emitted by neutron stars (NSs) and black holes (BHs) in a binary system~\cite{Chatziioannou:2014coa}.

%\3 Katerina's paper
		
The ability to break degeneracies among the binary's parameters and modifications to GR makes spin-precessing binaries ideal systems to probe parity violation~\cite{Alexander:2017jmt}. We here present the first ever study of precessing, quasi-circular compact binaries in a modified theory of gravity. We focus our attention on binary systems composed of spinning black holes, which, in dCS gravity, are modified from GR through scalar dipole hair and a perturbed quadrupole moment. We focus on the binary's evolution and GW emission during the inspiral phase of coalescence, where gravitational fields are weak and the orbital velocity is small compared to the speed of light. We thus work in the post-Newtonian (PN) formalism, and numerically construct, for the first time, the GWs emitted by spin-precessing, quasi-circular binaries in dCS gravity, including the effects of amplitude and frequency modulation due to spin-precession. Using a mismatch argument, we then estimate the level to which the dCS coupling parameter $\xi$ could be constrained with future GW observations.
	
\vspace{0.3cm}
\noindent \textit{Precession Equations.} When the binary is widely separated, the orbital time scale (i.e.~the orbital period) is much longer than the precession timescale (i.e.~the time it takes for a precession cycle to complete), which is in turn much longer than the radiation-reaction timescale (i.e.~the time it takes for the orbit to decay significantly). This separation of timescales allows us to consider the precession of angular momenta without radiation-reaction as a first approximation within PN theory~\cite{Loutrel-dCS}. 

The motion of spinning BHs has been well studied in the context of effective field theory within GR~\cite{Steinhoff:2010zz}. To obtain the precession equations for BHs in dCS gravity, we apply the same effective field theory methods by requiring the matter action to satisfy a set of additional symmetries, i.e. parity and shift invariance. We construct an action that contains all possible terms that satisfy these symmetries, and then restrict our attention to the subset that do not vanish when matching to the case of isolated BHs in dCS gravity.

%With this in mind, 
The variation of the action in effective field theory with respect to the linear and angular velocities of the BHs gives us the equations of motion for the linear and spin angular momenta. After solving for all fields within the near zone of the binary, we PN expand the spin-precession equations, orbit-average them and obtain
\begin{align}
\label{eq:S-dot}
\dot{\vec{S}}_{1} &= v^{5} \vec{\Omega}_{\text{\tiny SO}}^{\text{\tiny GR}} \times \vec{S}_{1} + v^{6} \left[\left(1 + \frac{25}{16} \frac{\xi}{m_{1}^{2} m_{2}^{2}}\right) \vec{\Omega}_{\text{\tiny SS}}^{\text{\tiny GR}}
\right.
\nn \\
&\left.
+ \left(1 - \frac{201}{112} \frac{\xi}{m_{1}^{4}}\right) \vec{\Omega}_{\text{\tiny MQ}}^{\text{\tiny GR}}\right] \times \vec{S}_{1}\,,
\end{align}
where $[\vec{\Omega}_{\text{\tiny SO}}^{\text{\tiny GR}}, \vec{\Omega}_{\text{\tiny SS}}^{\text{\tiny GR}}, \vec{\Omega}_{\text{\tiny MQ}}^{\text{\tiny GR}}]$ are the coefficients of the leading PN order spin-orbit, spin-spin, and monopole-quadrupole interactions in the GR spin-precession equations, specifically,
\begin{align}
\vec{\Omega}_{\text{\tiny SO}}^{\text{\tiny GR}} &= \mu \left(2 + \frac{3}{2} \frac{m_{2}}{m_{1}}\right) \hat{L}
\\
\vec{\Omega}_{\text{\tiny SS}}^{\text{\tiny GR}} &= \frac{1}{2 M^{3}} \left[\vec{S}_{2} - 3 \left(\hat{L} \cdot \vec{S}_{2}\right) \hat{L}\right]
\\
\vec{\Omega}_{\text{\tiny MQ}}^{\text{\tiny GR}} &= - \frac{3}{2} \frac{m_{2}}{m_{1}} \frac{1}{M^{3}} \left(\hat{L} \cdot \vec{S}_{1}\right) \hat{L}
\end{align}
with $M = m_{1} + m_{2}$ and $\mu = m_{1} m_{2} / M$ the total and reduced masses of a binary with component masses $m_{1}$ and $m_{2}$, spin angular momenta $\vec{S}_{1}$ and $\vec{S}_{2}$, orbital velocity $v$ and orbital angular momentum oriented along $\hat{L}$.
 
As we have found, in dCS gravity only the spin-spin and monopole-quadrupole interactions are modified. The latter of these is due to the fact that BHs in dCS gravity have a modified quadrupole moment. The former is unique since it arises due to a dipole-dipole interaction, when the scalar dipole moment of the first body interacts with the scalar dipole moment of the second body through the dCS pseudo-scalar field. Since the dipole moment is proportional to spin, this interaction can be recast as a spin-spin interaction. To obtain the precession equation for the spin of body 2, one can simply replace $1 \leftrightarrow 2$ in the above equation, and the precession equation for the orbital angular momentum is $\dot{\vec{L}} = -\dot{\vec{S}}_{1} - \dot{\vec{S}}_{2}$.

Before we study the GW emission from a BH binary in dCS gravity, it is useful to study the properties of the spin-precession equations in the absence of radiation reaction. The precession equations admit seven conserved quantities. Three of these are the magnitudes of the spin and orbital angular momenta $(|\vec{L}|, |\vec{S}_{1}|, |\vec{S}_{2}|)$, which are guaranteed by symmetry since there is no GW emission to carry angular momentum away from the binary or into the BHs' horizons. Three additional conserved quantities, specifically the components of the total angular momentum $\vec{J} = \vec{L} + \vec{S}_{1} + \vec{S}_{2}$, guarantee the existence of a co-precessing reference frame~\cite{Kesden:2014sla}. The remaining constant of motion is revealed upon inclusion of the quadrupole-monopole interactions in the precession equations~\cite{Racine:2008qv, Kesden:2014sla}. This seventh constant, referred to as the effective mass-weighted spin, represents the projection of the spin angular momenta of the bodies onto the direction of the orbital angular momentum. Unlike the other constants of motion, the effective mass-weighted spin is modified in dCS gravity due to the extra dipole-dipole and quadrupole-monopole interactions and we find that it is given by
\begin{align}
\Xi &= \left(1 + \frac{m_{2}}{m_{1}}\right) \left(1 + \frac{\xi}{m_{1}^{4}} {\cal{A}} + \frac{\xi}{m_{1}^{2} m_{2}^{2}} {\cal{C}}\right) \; \hat{L}\vec{S}_{1}
\nn \\
&+ \left(1 + \frac{m_{1}}{m_{2}}\right) \left(1 + \frac{\xi}{m_{2}^{4}} {\cal{B}} + \frac{\xi}{m_{1}^{2} m_{2}^{2}} {\cal{C}}\right) \; \hat{L}\vec{S}_{2}\,,
\end{align}
with
\allowdisplaybreaks
\begin{align}
{\cal{A}} &= \frac{201}{224} \frac{m_{2} \; \hat{L}\vec{S}_{1}}{\left[M |\vec{L}| - m_{2} \; \hat{L}\vec{S}_{1} - m_{1} \; \hat{L}\vec{S}_{2}\right]}\,,
\\
{\cal{B}} &= - \frac{201}{224} \frac{m_{2} \; \hat{L}\vec{S}_{1}}{m_{1} \; \hat{L}\vec{S}_{2}} \frac{\left[m_{2} \; \hat{L}\vec{S}_{1} + 2 m_{1} \; \hat{L}\vec{S}_{2}\right]}{\left[M |\vec{L}| - m_{2} \; \hat{L}\vec{S}_{1} - m_{1} \; \hat{L}\vec{S}_{2}\right]}\,,
\\
{\cal{C}} &= \frac{25}{48} \frac{m_{2} \; \hat{L}\vec{S}_{1}}{\left[m_{2} \; \hat{L}\vec{S}_{1} + m_{1} \; \hat{L}\vec{S}_{2}\right]} 
\frac{\left(m_{2} - m_{1}\right) |\vec{L}| - 3 m_{1} \; \hat{L}\vec{S}_{2}}{\left[M |\vec{L}| - m_{2} \; \hat{L}\vec{S}_{1} - m_{1} \; \hat{L}\vec{S}_{2}\right]}\,.
\end{align}
where we have introduced the short-hand $\hat{L}\vec{S}_{A} := \left(\hat{L} \cdot \vec{S}_{A}\right)$. The existence of these seven constants of motion plays a critical role in the construction of analytic waveforms for spin-precessing binaries, since they allow the reduction of the precession equations to quadratures. Such a calculation has already been performed in GR~\cite{Racine:2008qv, Kesden:2014sla}, with an extension to include radiation reaction in~\cite{Chatziioannou:2017tdw}, and the calculation of analytic, Fourier domain waveforms in~\cite{Chatziioannou:2016ezg}. The results obtained here would allow for the construction of similar spin-precessions waveforms in dCS gravity. 

\vspace{0.3cm}
\noindent \textit{Waveforms.} The derivation of an analytic Fourier domain waveform for spin-precessing binaries in dCS gravity is the ultimate goal for learning how to constrain deviations from GR in such systems. However, the derivation of such a waveform is exceedingly lengthy and goes well beyond the scope of this work. We will here consider a simplified, semi-analytic, waveform model~\cite{Lang:1900bz} which has already been used to study constraints on the graviton mass from GW observations of precessing binaries~\cite{Stavridis:2009mb}. We are here interested in how well we may be able to constrain dCS gravity with precessing binaries with GW observations, so we will adapt the waveforms specifically to aLIGO detectors.

The amplitude of the waveform is given by Eq.~(2.3) in~\cite{Stavridis:2009mb}, which is characterized by the directions of the orbital angular momentum and line of sight to the source, and the beam pattern functions $F_{+}$ and $F_{\times}$ of the detector. Since we focus on GW sources relevant to ground-based detectors, the beam pattern functions are still given by Eq.~(2.4) in~\cite{Stavridis:2009mb}, with the time-varying polarization angle still given by Eq.~(2.5) therein. For ground-based detectors, the amplitude must be rescaled by a factor of $2/\sqrt{3}$, although this is an overall scaling factor that is not relevant in this paper.

The Fourier phase of the waveform differs from~\cite{Stavridis:2009mb} due to the specific modification from dCS gravity. Following~\cite{Lang:1900bz, Stavridis:2009mb}, we write the Fourier phase as $\Phi(f) = \Psi(f) + \varphi_{\rm pol}[t(f)] + \delta_{\rm p}\Phi[t(f)]$, where $\Psi(f)$ is the contribution coming from the orbital phase that exists even in the non-precessing case and it can be found through the stationary phase approximation, $\varphi_{\rm pol}(f)$ is the polarization phase given in Eq.~(2.17) of~\cite{Stavridis:2009mb}, and $\delta_{\rm p}\Phi(f)$ is the integrated phase caused by precession of the orbital angular momentum and given in Eq.~(2.19) of~\cite{Stavridis:2009mb}. All of these Fourier phase contributions depend on the orbital or the spin angular momentum, and thus, they will acquire modifications due to dCS changes in the spin-precession equations found here.

Consider the orbital part of the Fourier phase $\Psi(f)$. In dCS gravity, the frequency evolution of the GWs was first found in~\cite{Yagi:2012vf} and it is modified due to the presence of dipole radiation: $\dot{f} = \dot{f}^{\rm GR} (1 + \delta C \eta^{-4/5} v^{4})$, where $\eta = \mu/M$ is the symmetric mass ratio, $\dot{f}^{\rm GR} = (96/5 \pi {\cal{M}}^{2}) v^{11}$ is the leading PN order, frequency chirping rate of GR, $v = (\pi {\cal{M}} f)^{1/3}$, with ${\cal{M}} = M \eta^{3/5}$ the chirp mass, and $\delta C$ given in Eq.~(9) of~\cite{Yagi:2012vf}\footnote{The numerical factors in our expression for $\delta C$ differ slightly from those in \cite{Yagi:2012vf} because of a different definition of the coupling constant $\xi$. The mapping from our definition of $\xi$ to theirs is $\xi \rightarrow \xi/16$.}. For spin-precessing binaries, $\delta C$ becomes oscillatory, and the inversion to obtain $t(f)$ and $\Psi(f)$ would in general be non-trivial. However, as shown in~\cite{Cutler:1994ys}, such oscillations are also present in GR at 2PN order and they lead to small corrections in the inversion that can be neglected without a significant loss of accuracy. We have verified that the effect of the oscillations of $\delta C$ in the inversion is also small, and that subsequently, we can treat $\delta C$ as a constant in the inversion. This implies that $\Psi(f) = 2\pi f t_{c} - \phi_{c} - \pi/4 + (3/128)v^{-5}(1 - 10 \delta C \eta^{-4/5} v^{4})$ and $t(f) = t_{c} - (5/256) {\cal{M}} v^{-8} (1 - 2 \delta C \eta^{-4/5} v^{4})$, with $t_{c}$ and $\phi_{c}$ the time and phase of coalescence, and $\delta C = \delta C[t(f)]$.

The remaining two contributions to the Fourier phase, $\varphi_{\rm pol}$ and $\delta_{\rm p} \Phi$, are still given by Eqs.~(2.17) and (2.19) in~\cite{Stavridis:2009mb}, respectively, but they no longer have the same time evolution as in GR, since they are now governed by the dCS precession equations in Eq.~\eqref{eq:S-dot}. Thus, to obtain $\varphi_{\rm pol}(t)$ and $\delta_{\rm p} \Phi(t)$, we numerically solve the dCS spin-precession equations to obtain $\hat{L}(t)$. We then calculate $\varphi_{\rm pol}(t)$ and $\delta_{\rm p}\Phi(t)$ numerically and finally convert them into functions of frequency using the inversion for $t(f)$. 

\begin{figure}[ht]
\includegraphics[clip=true,width=\columnwidth]{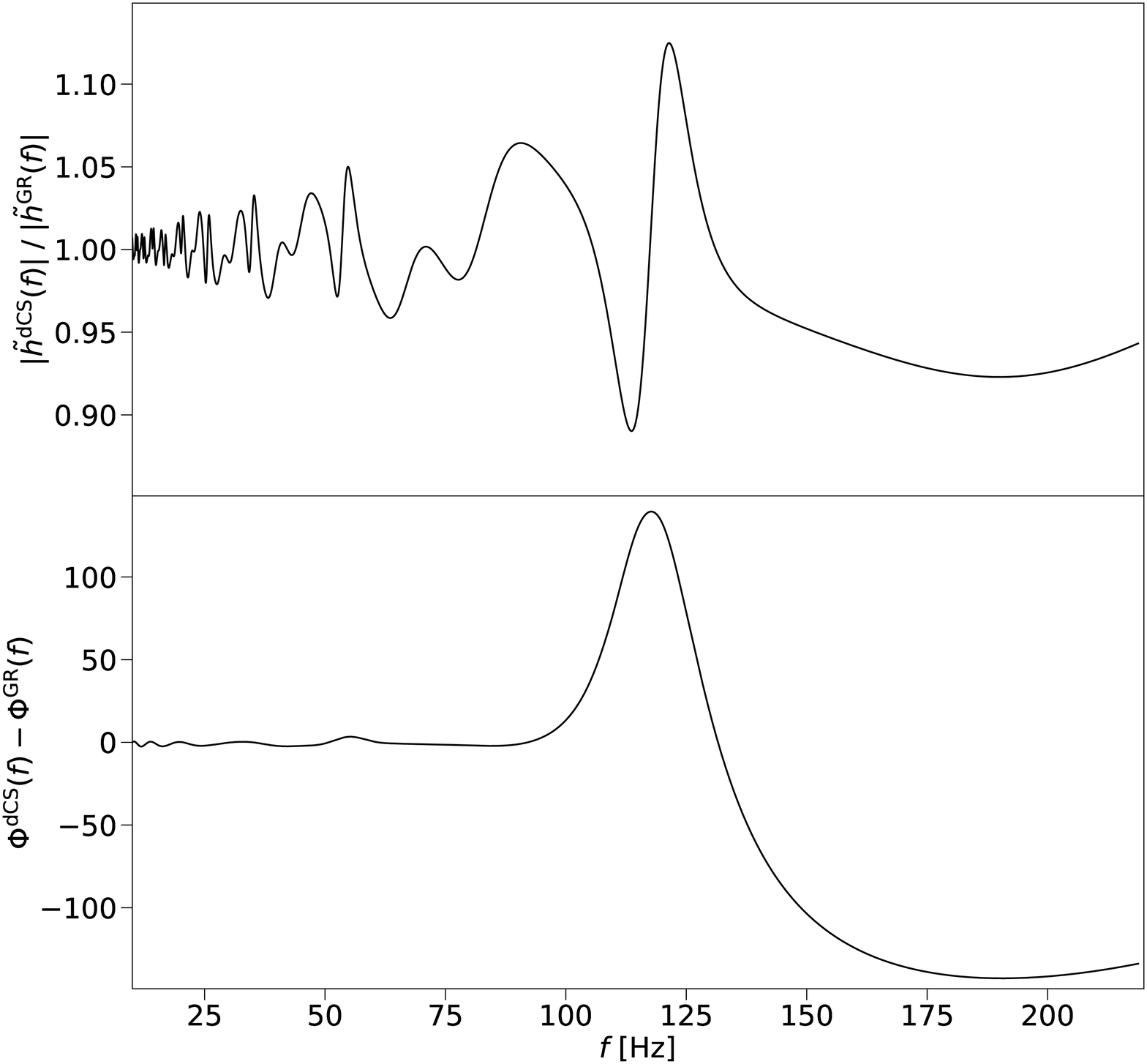}
\caption{\label{compare} Normalized amplitude (top) and dephasing (bottom) of the waveform for a spin-precessing binary in dCS gravity relative to GR.}
\end{figure}

Figure~\ref{compare} shows the differences between spin-precessing waveforms in GR and in dCS gravity. As a representative example, we choose an equal-mass system with $|\vec{S}_{1}|/m_{1}^{2} := \chi_{1} = 0.9$ and $|\vec{S}_{2}|/m_{2}^{2} := \chi_{2} = 0.95$, $\hat{S}_{1} = (0.9, 0, 0.44)$ and $\hat{S}_{2} = (0.1, 0.25, 0.96)$, and $m_{1} = 10 M_{\odot} = m_{2}$. We numerically evolve the binary starting at $f_{\rm GW} = 2/T_{\rm orb} = 5 \; {\rm Hz}$, where $T_{\rm orb}$ is the orbital period of the binary, and stop the numerical evolution when the system reaches the last stable orbit for a Schwarzschild BH, $v = 1/\sqrt{6}$. For the dCS case, we take $\xi^{1/4} = 10 \; {\rm km}$, which we have verified satisfies the weak-coupling approximation required for the effective field theory treatment to remain valid.
The top panel shows the amplitude of the dCS waveform normalized to the amplitude of the GR waveform, while the bottom panel shows the difference in the total Fourier phase between the dCS and GR waveforms. The amplitude displays modulations typical of spin-precessing systems, but these modulation are modified by the dCS effects. Of the Fourier phase components, the precession component $\delta_{\rm p}\Phi$ displays the largest deviations from GR, and dominates the dephasing between the dCS and GR waveforms. This is to be expected since the modifications to the spin-precession equations enter at leading PN order, while corrections to the orbital part of the Fourier phase enter at 2PN order.

\vspace{0.3cm}
\noindent {\textit{Parameter Estimation.}} We now study the degree to which we can constrain dCS gravity with spin-precessing GW observations. To obtain an estimate of the projected constraints on $\xi$, we consider a match argument, which involves determining the faithfulness of a GR waveform template in recovering a dCS waveform. We begin by defining the noise-weighted inner product between two Fourier domain waveforms $\tilde{h}_{1}$ and $\tilde{h}_{2}$ as
\begin{equation}
\Big(\tilde{h}_{1} \; \Big| \; \tilde{h}_{2}\Big) = 2 \int_{f_{\rm min}}^{f_{\rm max}} df \frac{\tilde{h}_{1}(f) \tilde{h}_{2}^{*}(f) + \tilde{h}_{1}^{*}(f) \tilde{h}_{2}(f)}{S_{n}(f)}\,,
\end{equation}
where $^{*}$ corresponds to complex conjugation and $S_{n}(f)$ is the power spectral density of the GW detector. We are here interested in observations with aLIGO, so we use the $S_{n}(f)$ for aLIGO at design sensitivity~\cite{Matt-priv}, and we take $f_{\rm min} = 10 \; {\rm Hz}$ and $f_{\rm max} = f_{\rm merger}$. The match between two waveforms is then given by 
\begin{equation}
\text{M} = \underset{\delta \phi, \delta t}{\rm max} \; \frac{\Big(\tilde{h}_{1} \; \Big| \; \tilde{h}_{2} e^{i (\delta \phi + 2\pi f \delta t)}\Big)}{\sqrt{\Big(\tilde{h}_{1} \; \Big| \; \tilde{h}_{1}\Big) \Big(\tilde{h}_{2} \; \Big| \; \tilde{h}_{2}\Big)}}\,,
\end{equation}
where $\delta \phi$ and $\delta t$ are overall phase and time shifts that one maximizes the inner product over. 

The match provides us a means of determining the minimum value of $\xi$ such that mismodeling, specifically not including dCS effects in the precessing waveform, does not significantly bias parameter estimation. This can be translated into requiring the systematic error due to mismodeling be smaller than statistical measurement error. The nominal match that satisfies this threshold is $\text{M} = 1 - D/(2 \; {\rm SNR}^{2})$, where $D$ is the number of parameters in the waveform and ${\rm SNR} = (h | h)^{1/2}$ is the signal-to-noise ratio~\cite{Chatziioannou:2017tdw}. For the systems considered here, $D=15$ and for a binary with ${\rm SNR} = 100 (50) [25]$ this corresponds to a nominal threshold of $\text{M} = 0.9993 (0.997) [0.99]$.

\begin{figure}[ht]
\includegraphics[clip=true,width=\columnwidth]{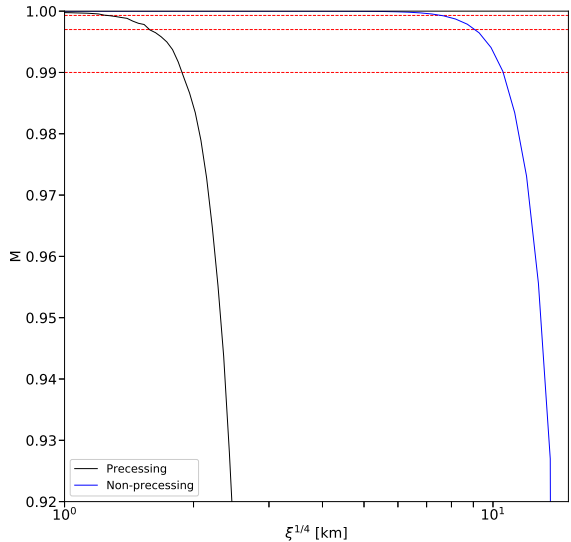}
\caption{\label{match} Match between GR and dCS waveforms for a spin-precessing binary as a function of $\xi^{1/4}$ for a precessing system (black line) and a non-precessing system (blue line). The red dashed lines corresponds to the nominal matches $\text{M} = 0.9993, 0.997, 0.99$ at ${\rm SNR} = 100, 50, 25$, repsectively.}
\end{figure}

To estimate projected constraints on $\xi$, we study the representative system discussed earlier, and compute the match for various values of $\xi$, as shown in Fig.~\ref{match}. For comparison, we consider a system with the same masses, spins, and initial separation, but which is non-precessing, i.e. where the spins and orbital angular momentum are aligned. Using the nominal match described above, $\xi^{1/4} \lesssim 1.3 (1.6) [1.9] \; {\rm km}$ for the precessing system, while $\xi^{1/4} \lesssim 7.6 (9.2) [10.5] \; {\rm km}$ for the non-precessing system, to ensure that the systematic error is less than the statistical error. The precessing binary allows for roughly an order of magnitude better constraint than the non-precessing system, and an eight order of magnitude better constraint than current Solar System experiments.

The estimate above does not account for possible covariances between the parameters of the binary and the dCS modification, which may somewhat deteriorate our ability to constrain $\xi$. Further, the scalar dipole moment of BHs is known to all orders of spin. For certain values of the scalar dipole moment, dipole contributions to the GW energy flux can vanish~\cite{Yagi:2012vf}, which would impact our ability to place constraints on $\xi$. This constraint, however, would also improve inversely with the signal-to-noise ratio. A more detailed Bayesian analysis can be carried out, once analytic Fourier-domain waveforms are calculated for dCS-corrected, spin-precessing binaries. The work presented here allows not only for such an analysis, but also for the construction of phenomenological inspiral-merger-ringdown waveforms that can be compared directly to GW observations in the future.   

 %-----------------------------------------------------
\acknowledgments 
N. L. and N. Y. acknowledge support from NSF EAPSI Award No. 1614203, NSF CAREER grant PHY-1250636, and NASA grants NNX16AB98G and 80NSSC17M0041. T. T. acknowledges support in part by MEXT Grant-in-Aid for Scientific Research on Innovative Areas, Nos. 17H06357 and 17H06358, and by Grant-in-Aid for Scientific Research Nos. 26287044 and 15H02087. 

%%%%%%%%%%%%%%%%%%%%%%%%%%%%%%%%%%%%%%%%%%%%
%%%%%%%%%%%%%%%%%%%%%%%%%%%%%%%%%%%%%%%%%%%%
\bibliography{master}
\end{document}